\documentclass[12pt]{article}

\usepackage{amsmath}
\usepackage{graphicx}
\usepackage{eufrak}

\renewcommand{\theequation}{\arabic{section}.\arabic{equation}}
\newcommand{\mys}[1]{\section{#1} \setcounter{equation}{0}}

\newcommand{\myappendix}{\appendix
   \renewcommand{\theequation}{\Alph{section}.\arabic{equation}}
   \vspace{30pt} \noindent {\Large \bf Appendix}}

\newlength{\dummysp}
\newcommand{\diag}{\mathop{{\hbox{diag} \, }}\nolimits}
\newcommand{\tr}{\mathop{{\hbox{Tr} \, }}\nolimits}

\newcommand{\stxt}[1]{\mathop{\hbox{{\scriptsize #1}}}\nolimits}
\newcommand{\bbar}[1]{{\overline{#1}}}
\newcommand{\half}{\frac{1}{2}}

\newcommand{\beq}{\begin{eqnarray}}
\newcommand{\eeq}{\end{eqnarray}}
\newcommand{\nnn}{ \nonumber \\ }

\newcommand{\Rbf}{{{\bf R}}}

\newcommand{\e}{{\epsilon}}
\newcommand{\s}{{\sigma}}
\newcommand{\vev}[1]{{\langle #1 \rangle}}
\newcommand{\bigvev}[1]{{\left\langle #1 \right\rangle}}
\newcommand{\ord}[1]{{{\cal O}(#1)}}
\newcommand{\gappeq}{\mathrel{\rlap {\raise.5ex\hbox{$>$}}
{\lower.5ex\hbox{$\sim$}}}}
\newcommand{\lappeq}{\mathrel{\rlap{\raise.5ex\hbox{$<$}}
{\lower.5ex\hbox{$\sim$}}}}
\newcommand{\myref}[1]{(\ref{#1})}

\newcommand{\ben}{\begin{enumerate}}
\newcommand{\een}{\end{enumerate}}

\newcommand{\sqtw}{\sqrt{2}}
\newcommand{\fourth}{\frac{1}{4}}

\newcommand{\psib}{{\bar \psi}}

\newcommand{\bit}{\begin{itemize}}
\newcommand{\eit}{\end{itemize}}

\newcommand{\Cbf}{{\bf C}}

\newcommand{\obf}{{\bf 1}}
\newcommand{\nbf}{{\bf n}}
\newcommand{\mbf}{{\bf m}}
\newcommand{\kbf}{{\bf k}}
\newcommand{\xb}{{\bbar{x}}}

\newcommand{\ibf}{\boldsymbol{\hat \imath}}
\newcommand{\jbf}{\boldsymbol{\hat \jmath}}
\newcommand{\lbf}{\boldsymbol{\ell}}
\newcommand{\sbf}{\boldsymbol{\s}}

\newcommand{\Ncal}{{\cal N}}

\newcommand{\SSB}{{S_{\stxt{SB}}}}
\newcommand{\hphi}{{\hat \phi}}
\newcommand{\hg}{{\hat g}}
\newcommand{\hmu}{{\hat \mu}}
\newcommand{\hginv}{{{\hat g}^{-1}}}
\newcommand{\muinv}{{{\hat \mu}^{-1}}}
\newcommand{\zb}{{\bbar{z}}}
\newcommand{\xdag}{x^\dagger}
\newcommand{\ydag}{y^\dagger}
\newcommand{\zdag}{z^\dagger}
\newcommand{\hx}{{\hat x}}
\newcommand{\hxd}{{\hat x}^\dagger}
\newcommand{\hxb}{{\hat \xb}}
\newcommand{\hy}{{\hat y}}

\newcommand{\hz}{{\hat z}}
\newcommand{\Dslash}{{\not \hspace{-4pt} D} }

\def\[{\left [}
\def\]{\right ]}
\def\({\left (}
\def\){\right )}

\begin{document}

\begin{titlepage}

\renewcommand{\thefootnote}{\fnsymbol{footnote}}

\hfill Dec.~19, 2003

\hfill hep-lat/0312020

\vspace{0.45in}

\begin{center}
{\bf \Large Deconstruction, 2d lattice Yang-Mills, 
\\ \vskip 10pt and the dynamical lattice spacing}
\end{center}

\vspace{0.15in}

\begin{center}
{\bf \large Joel Giedt\footnote{{\tt giedt@physics.utoronto.ca}}}
\end{center}

\vspace{0.15in}

\begin{center}
{\it University of Toronto \\
60 St. George St., Toronto ON M5S 1A7 Canada}
\end{center}

\vspace{0.15in}

\begin{abstract}
We study expectation values related to the
dynamical lattice spacing that occurs in the recent
supersymmetric 2d lattice Yang-Mills constructions
of Cohen et al.~[hep-lat/0307012].  For the purposes
of this preliminary analysis, we restrict our attention 
to the bosonic part of that theory.
That is, we compute observables in the fully quenched
ensemble, equivalent to non-supersymmetric 2d lattice Yang-Mills
with a dynamical lattice spacing and adjoint scalars.  
Our numerical simulations 
indicate difficulties with the proposed continuum limit.
We find that expectation values tend to those
of the undeformed ``daughter theory,''
in spite of the deformation suggested by Cohen et al.
In an effort to understand these results, we
examine the zero action configurations, with and without
the deformation.  Based on these considerations,
we are able to interpret the simulation
results in terms of entropic effects.

\end{abstract}

\end{titlepage}

\renewcommand{\thefootnote}{\arabic{footnote}}
\setcounter{footnote}{0}

\mys{Introduction}
In \cite{Cohen:2003qw}, a lattice action has been proposed
by Cohen, Kaplan, Katz and Unsal (CKKU) for the
(4,4) 2d $U(k)$ super-Yang-Mills.\footnote{The
(4,4) 2d $U(k)$ super-Yang-Mills is best defined
as the dimensional reduction of $\Ncal=1$ 6d $U(k)$
super-Yang-Mills; the notation ``(4,4)'' denotes
the number of left and right 2d chirality supercharges.}
The Euclidean target theory action is:
\beq
S_{(4,4)} &=& \int d^2 x \; \frac{1}{g_2^2} \tr \[ (Ds_\mu) \cdot (Ds_\mu)
+ \psib_i \Dslash \psi_i + \fourth F \cdot F \right. \nnn
&& \left. 
+ \psib_i [ (s_0 \delta_{ij} + i \gamma_3 {\bf s} \cdot \sbf_{ij}) ,\psi_j] 
- \half [s_\mu, s_\nu]^2 \]
\label{wkkr}
\eeq
where $s_\mu$ ($\mu=0,1,2,3$) are hermitian scalars, $F$ is
the 2d Yang-Mills field strength, and $\psi_i$ ($i=1,2$) are
2d Dirac fermions, all in the adjoint representation
of $U(k)$.  In \cite{Giedt:2003vy}, some aspects
of the fermion determinant were examined, with results
very similar to those in \cite{Giedt:2003ve}.  Here we will
study a rather fundamental aspect of the construction:
the proposed emergence of a ``dynamical'' lattice spacing.
CKKU make use of an $N \times N$ lattice that contains
link and site fields that are $k \times k$ matrices.
The classical lattice action contains many zero action
configurations.  CKKU expand the classical lattice action 
about a particular class of configurations (discussed
below) that are characterized by a parameter $a$.  For this
reason we will refer to such a configuration as an
{\it $a$-configuration.}  In the limit $a \to 0, \; N \to \infty$,
the classical action tends to the continuum action
of (4,4) 2d $U(k)$ super-Yang-Mills.
Thus $a$ is interpreted as a lattice spacing.  
However, it is dynamical as it has to
do with a particular background configuration for the lattice
fields.  This strategy is based on the ideas of {\it deconstruction}
\cite{Arkani-Hamed:2001ca,Hill:2000mu}.  Studies
where a partial latticization of 4d supersymmetric theories
has been obtained by this approach include \cite{Giedt:2003xr,Poppitz:2003uz}.

It is obvious that
to arrive at the proposed continuum theory requires that
a semi-classical expansion about the $a$-configuration 
gives a good approximation to the behavior of the 
full lattice theory.  But all $a$-configurations are
energetically equivalent for any value of $a$.  Furthermore,
there exist other zero action configurations that do
not fall into the $a$-configuration class (shown below).  For this
reason, CKKU deform the action by adding an $a$-dependent 
potential that favors the $a$-configuration.  The ``continuum
limit'' then includes sending $a \to 0$ in this potential.
Although this deformation breaks the exact lattice supersymmetry,
it is rendered harmless by scaling the relative strength of
the deformation potential to zero in the thermodynamic limit.
For this reason it has been argued by CKKU that the quantum
continuum limit is nothing but the target theory, without
the need for fine-tuning.
For further details, we refer the reader to \cite{Cohen:2003qw},
as well as the articles leading up to 
it \cite{Kaplan:2002wv,Cohen:2003xe}.

The deconstruction method for latticizing a 2d continuum target theory
does not require fermions or supersymmetry.
The bosonic system (i.e., setting all lattice fermions
to zero) contained in the model of
CKKU already has the interesting feature of a deconstructed
lattice Yang-Mills.  The same semi-classical 
arguments yield $U(k)$ Yang-Mills with 4
adjoint scalars; that is, the bosonic part of \myref{wkkr}.
In the present article we will investigate the validity
of the semi-classical argument.  We will take into account 
quantum effects, estimating key expectation values
by means of Monte Carlo simulation.  Of course
we expect such results to differ from those that would
be obtained in the related supersymmetric theory of CKKU;
for in that case one would have dynamical fermions,
and in the limit of vanishing deformation, supersymmetric
nonrenormalization theorems.  However, we find it sufficiently
interesting to analyze the bosonic system as a preliminary
step.  Indeed, for the expectation values that
we study, we find that semi-classical arguments
are not a good indicator of the full quantum behavior.

In the next section we introduce the essential features of
the bosonic part of the CKKU construction that will be needed
for the subsequent discussion.  In Section \ref{s:cla} we study the
classical minima of the undeformed and deformed actions.
In Section \ref{s:sim} we outline the methods and results of
our lattice simulation.  In Section \ref{intp} we suggest an interpretation
of these results in terms of the analysis of Section \ref{s:cla}.
In Section \ref{s:con} we make some concluding remarks.  In
the Appendix, we provide a brief review of the $\Ncal=4$
4d super-Yang-Mills moduli space, which arises in the
discussion of Section \ref{s:cla}.

\mys{Quiver lattice Yang-Mills}
For CKKU, the starting point is the Euclidean
$\Ncal=1$ 6d $U(kN^2)$ super-Yang-Mills.
In our case, the starting point will instead be 
Euclidean 6d $U(kN^2)$ Yang-Mills.
This is dimensionally reduced to 0d to obtain a $U(kN^2)$ matrix model.
The matrix model naturally possesses $SO(6)$ Euclidean invariance.
Next we note $SO(6) \supset SO(2) \times SO(2) \supset 
Z_N \times Z_N$.  CKKU have identified a homomorphic embedding 
of $Z_N \times Z_N$ into the $U(kN^2)$ gauge symmetry group.
Using this, a $Z_N \times Z_N$ orbifold projection is performed to
obtain a $U(k)^{N^2}$ 0d {\it quiver,} or, {\it product group} 
theory.\footnote{Quiver theories were originally studied
many years ago in other contexts \cite{Georgi:au,Halpern:1975yj}.}  
In every respect we follow CKKU, except that we have
set all fermions to zero.  For details of the matrix model and
orbifold projection, we refer the reader to \cite{Cohen:2003qw};
in the interests of brevity, we will only give the final 
result, the {\it undeformed} lattice action:
\beq
S_0 &=& \frac{1}{g^2} \tr \sum_{\nbf} \[
\half (\xdag_{\nbf-\ibf} x_{\nbf-\ibf} - x_{\nbf} \xdag_{\nbf}
+ \ydag_{\nbf-\jbf} y_{\nbf-\jbf} - y_\nbf \ydag_\nbf
+ \zdag_\nbf z_\nbf - z_\nbf \zdag_\nbf )^2 \right. \nnn
&& + 2(x_\nbf y_{\nbf+\ibf} - y_\nbf x_{\nbf+\jbf})
(\ydag_{\nbf+\ibf} \xdag_\nbf - \xdag_{\nbf+\jbf} \ydag_\nbf) \nnn
&& + 2(y_\nbf z_{\nbf+\jbf} - z_\nbf y_\nbf)
(\zdag_{\nbf+\jbf} \ydag_\nbf - \ydag_\nbf \zdag_\nbf) \nnn 
&& \left. + 2(z_\nbf x_\nbf - x_\nbf z_{\nbf+\ibf})
(\xdag_\nbf \zdag_\nbf - \zdag_{\nbf+\ibf} \xdag_\nbf) \]
\label{sbta}
\eeq
Here, $x_\mbf, y_\mbf, z_\mbf$ are bosonic lattice
fields that are $k \times k$ unconstrained
complex matrices; $\mbf=(m_1,m_2)$ labels points on an $N \times N$
lattice, and $\ibf=(1,0), \jbf=(0,1)$ are unit vectors.
The $U(k)^{N^2}$ symmetry is nothing but the local $U(k)$
symmetry of the lattice action $S_0$, with link bosons $x_\mbf$ in
the $\ibf$ direction, link bosons $y_\mbf$ in the $\jbf$ direction, and
sites bosons $z_\mbf$, all transforming in the usual manner:
\beq
x_\mbf \to \alpha_\mbf x_\mbf \alpha_{\mbf + \ibf}^\dagger, \qquad
y_\mbf \to \alpha_\mbf y_\mbf \alpha_{\mbf + \jbf}^\dagger, \qquad
z_\mbf \to \alpha_\mbf z_\mbf \alpha^\dagger_\mbf
\label{yuer}
\eeq
Canonical mass dimension 1 is assigned to $x_\mbf, y_\mbf, z_\mbf$,
whereas $g$ has mass dimension 2.

Although $S_0$ is a lattice action that describes a
statistical system with interesting features, it is not in
any obvious way related to a 2d continuum field theory.
As will be explained below, $S_0 \geq 0$ and 
a vast number of nontrivial solutions to $S_0=0$ exist, not all of
which are gauge equivalent.  In fact, the space of minimum
action configurations, or {\it moduli space,} is a multi-dimensional
noncompact manifold with various {\it branches} (classes of
configurations).  For this reason it is difficult to say what a
``continuum limit'' might be; for there exists an infinite number
of energetically equivalent configurations about which to
expand, not all of which are gauge equivalent.

A surprising result---pointed out by CKKU, and based on
ideas from deconstruction---is obtained if one 
expands about the {\it $a$-configuration}
\beq
x_\mbf = \frac{1}{a \sqtw} \obf, \quad
y_\mbf = \frac{1}{a \sqtw} \obf, \quad
z_\mbf = 0, \quad \forall \mbf
\label{ckcf}
\eeq
keeping $g_2=ga$ and $L=Na$ fixed, treating $a$ as small.
(It is easy to see that $S_0=0$ for this configuration.)
That is, we associate $a$ with a lattice spacing (mass
dimensions -1), even though it arises originally from
a specific background field configuration.
In this case, one finds that the classical continuum limit is nothing
but the bosonic part of \myref{wkkr},
which is a variety of 2d $U(k)$ Yang-Mills with adjoint scalars.
In the case of the supersymmetric quiver theory of CKKU,
where fermions are present, one obtains \myref{wkkr}
in full; i.e., (4,4) 2d $U(k)$ super-Yang-Mills.

The trick is how to make the configuration \myref{ckcf}
energetically preferred without destroying all of the
pleasing symmetry properties of the theory.  (This is
particularly true in the supersymmetric case.)  In the
quantum analysis, we must address the more delicate
complication of entropy as well.  CKKU suggest a
deformation of the bosonic action in an effort to stabilize
the theory near the $a$-configuration \myref{ckcf}:
\beq
S &=& S_0 + \SSB \\
\SSB &=& \frac{a^2 \mu^2}{2 g^2} \sum_\nbf \tr 
\[ \(x_\nbf \xdag_\nbf -\frac{1}{2a^2}\)^2
+ \(y_\nbf \ydag_\nbf-\frac{1}{2a^2}\)^2 
+ \frac{2}{a^2} z_\nbf \zdag_\nbf \] 
\eeq
Here the strength of the deformation is determined
by the quantity $\mu$, which has mass dimension 1.
It is clear that the configuration \myref{ckcf}
minimizes $\SSB$.  (Other configurations that
minimize $S_0$ and $\SSB$ will be discussed below.)
Unfortunately, in the supersymmetric version of CKKU,
the deformation $\SSB$ breaks the exact supersymmetry of their original
lattice action (hence the subscript ``SB'').  For this reason they demand that
the strength of $\SSB$ relative to $S_0$, conveyed by $\mu^2$,
be scaled to zero in the thermodynamic limit.
Thus we are mostly interested in the effects of
the deformation subject to this scaling.

In much of what follows we will specialize to the case
of $U(2)$.  This is merely because it is the simplest
case and the most efficient to simulate.  In this special
case, $x_\mbf, y_\mbf, z_\mbf$ will be unconstrained $2 \times 2$
complex matrices.

\mys{The classical analysis}
\label{s:cla}
Here details are given of the classical analysis of
the minima of the action.  In Section \ref{jirr} we
consider the undeformed action $S_0$; then in
Section \ref{wqfe} we consider the modifications induced
by the deformation $\SSB$, which has the effect of
lifting some directions in moduli space.
Through understanding this classical picture,
naive expections of what will occur in the quantum
theory, based on energetics, can be formulated.

\subsection{Undeformed theory}
\label{jirr}
Here we neglect $\SSB$ and examine the minima of $S_0$.
Note that \myref{sbta} is a sum of terms of the form
$\tr A A^\dagger$ (the first line involves squares of
hermitian matrices).  Thus $S_0 \geq 0$ with $S_0 = 0$
iff the following equations hold true:
\beq
&& \xdag_{\nbf-\ibf} x_{\nbf-\ibf} - x_{\nbf} \xdag_{\nbf}
+ \ydag_{\nbf-\jbf} y_{\nbf-\jbf} - y_\nbf \ydag_\nbf
+ [\zdag_\nbf, z_\nbf] = 0 
\label{wrse} \\
&& x_\nbf y_{\nbf+\ibf} - y_\nbf x_{\nbf+\jbf} = 
y_\nbf z_{\nbf+\jbf} - z_\nbf y_\nbf = 
z_\nbf x_\nbf - x_\nbf z_{\nbf+\ibf} = 0
\label{wrsf}
\eeq
together with the h.c. of \myref{wrsf}.  The set
of solutions is the moduli space of the undeformed theory.  

\subsubsection{Zeromode branch}
To begin a study of the moduli space,
we isolate the zero momentum modes:  $x_\nbf \equiv x \; \forall
\nbf$, etc.  Then Eqs.~\myref{wrse} and \myref{wrsf} reduce to
\beq
&& [\xdag,x] + [\ydag,y] + [\zdag,z] = 0 \nnn
&& [x,y] = [y,z] = [z,x] = 0
\label{qers}
\eeq
together with the h.c. of the second line of \myref{qers}.  Eqs.~\myref{qers}
may be recognized as nothing but the {\it D-flatness}
and {\it F-flatness} constraints that describe
the moduli space associated with the classical
scalar vacuum of $\Ncal=4$ 4d super-Yang-Mills.\footnote{We
thank Erich Poppitz for pointing this out to us,
as well as the branch of moduli space \myref{kerr} given below.}
The equations are invariant with respect to the global
gauge transformation
\beq
x \to \alpha x \alpha^\dagger, \quad
y \to \alpha y \alpha^\dagger, \quad
z \to \alpha z \alpha^\dagger
\label{oore}
\eeq
Then it is well-known that solutions to \myref{qers}
consist of $x,y,z$ that lie in a Cartan subalgebra
of $U(k)$; the proof is reviewed in Appendix A.  
The global gauge transformations \myref{oore}
allow one to change to a basis where this Cartan subalgebra
has a diagonal realization.  Thus one can think of the
moduli space as the set of all possible diagonal matrices $x,y,z$,
and all global gauge transformations \myref{oore} of this set.

In particular, the zeromode moduli space of the undeformed
$U(2)$ theory is completely described by
\beq
x = x^0 + x^3 \s^3, \quad 
y = y^0 + y^3 \s^3, \quad 
z = z^0 + z^3 \s^3,
\label{uier}
\eeq
with arbitrary complex numbers $x^0,x^3,y^0,y^3,z^0,z^3$,
together with $U(2)$ transformations of these solutions.

Eqs.~\myref{wrse} and \myref{wrsf} also have non-zeromode
solutions.  We do not attempt to present an exhaustive
account of them.  We will merely point out a few such branches
in order to illustrate that the undeformed theory
has a very complicated and large set of $S_0=0$ configurations.
This observation will be relevant to our interpretation
of the simulation results in Section \ref{intp}.
  
\subsubsection{$x_\mbf=y_\mbf=0$  non-zeromode branch}  
We have the very ``large'' branch of moduli space described by
\beq
x_\mbf=y_\mbf=0, \quad z_\mbf = z_\mbf^0 + z_\mbf^3 \s^3, \quad
\forall \mbf
\label{kerr}
\eeq
Again, $z_\mbf^0,z_\mbf^3$ are arbitrary complex numbers.
Furthermore, $z_\mbf$ is a site variable and thus transforms
independently at each site as
\beq
z_\mbf \to \alpha_\mbf z_\mbf \alpha^\dagger_\mbf
\eeq

It can be seen that this branch affords a vast number of solutions to
\myref{wrse} and \myref{wrsf}; there are $N^2$ such
solutions, modulo choices for $z_\mbf^0, z_\mbf^3 \in \Cbf$
and gauge equivalences.  We will argue below that an understanding of
the entropic effects that result from this branch is necessary
to understanding expectation values in the quantum theory---even when
a potential is introduced which gives these configurations
nonvanishing action.

\subsubsection{$z_\nbf =0$ non-zeromode branch}
Another branch in moduli space is the following.  First we
set $z_\nbf =0, \; \forall \nbf$, and introduce Fourier space
variables
\beq
x_\nbf = \frac{1}{N} \sum_\kbf \omega^{\kbf \cdot \nbf} f_\kbf,
\qquad
y_\nbf = \frac{1}{N} \sum_\kbf \omega^{\kbf \cdot \nbf} g_\kbf,
\qquad
\omega = \exp(2\pi i/N)
\eeq
where $\kbf=(k_1,k_2)$ and $k_1, k_2 \in [0,1,\ldots,N-1]$.
Then taking into account $z_\nbf=0$, the conditions \myref{wrse}
and \myref{wrsf} are equivalent to:
\beq
0 &=& \sum_\kbf \( \omega^{\ibf \cdot \lbf} f_\kbf^\dagger f_{\kbf-\lbf}
- f_\kbf f_{\kbf+\lbf}^\dagger
+ \omega^{\jbf \cdot \lbf} g_\kbf^\dagger g_{\kbf-\lbf}
- g_\kbf g_{\kbf+\lbf}^\dagger \) \nnn
0 &=& \sum_\kbf \( \omega^{-\ibf \cdot (\lbf + \kbf)} f_\kbf g_{-\kbf-\lbf}
- \omega^{\jbf \cdot \kbf} g_{-\kbf-\lbf} f_\kbf \)
\label{klsf}
\eeq
for all $\lbf=(\ell_1,\ell_2)$ and $\ell_1, \ell_2 \in [0,1,\ldots,N-1]$.
Next we turn off all modes except one for both $f_\kbf$ and $g_\kbf$:
\beq
f_\kbf = \delta_{\kbf,\kbf'} f_{\kbf'}, \qquad
g_\kbf = \delta_{\kbf,-\kbf'} g_{-\kbf'}
\eeq
Here and below, {\it no} sum over $\kbf'$ is implied.
When substituted into \myref{klsf}, only 2 nontrivial
conditions survive:
\beq
0 = [f_{\kbf'}^\dagger, f_{\kbf'}]
+ [g_{-\kbf'}^\dagger, g_{-\kbf'}], \qquad
0 = f_{\kbf'} g_{-\kbf'} 
- \omega^{(\ibf+\jbf) \cdot \kbf'} g_{-\kbf'} f_{\kbf'}
\label{klsg}
\eeq
For the $U(2)$ case, we find that solutions exist if
$\omega^{(\ibf+\jbf)\cdot \kbf'} = \pm 1$.  We already know from the
zeromode considerations that for $\omega^{(\ibf+\jbf)\cdot \kbf'} = 1$
we have solutions for $f_{\kbf'}, g_{-\kbf'}$
diagonal matrices.  In the case of $\omega^{(\ibf+\jbf)\cdot \kbf'} = -1$
it is easy to see that there are solutions, say, of the form
\beq
f_{\kbf'} = z_f \s^3, \quad g_{-\kbf'} = z_g (\s^1 + b \s^2),
\quad z_f, z_g \in \Cbf, \quad b \in \Rbf
\label{ppor}
\eeq
There are many values of $\kbf'$ for
which $\omega^{(\ibf+\jbf)\cdot \kbf'} = \pm 1$.  
For $N$ even these are
\beq
k'_1 + k'_2 = 0, \frac{N}{2}, N, \frac{3N}{2}
\eeq
For $N$ odd, $k'_1 + k'_2 = 0, N$ are allowed and in the cases
where
\beq
k'_1 + k'_2 = \frac{N \pm 1}{2}, \frac{3 (N\pm 1)}{2}
\eeq
\myref{ppor} yield approximate solutions to \myref{klsg}, with an error
of order $1/N$.  Thus in the $N \to \infty$ limit the number 
of $S_0=0$ configurations in this class is vast; in fact, it
is easy to check that the number of such configurations is
approximately $2N$, modulo gauge equivalences and
various choices for the constants in \myref{ppor}.

\subsection{Deformed theory}
\label{wqfe}
Now we consider the supersymmetry breaking
deformation $\SSB$ introduced by CKKU.  
To see its effect it is handy to rewrite
the quantities that appear in it.  Recall that $x_\mbf$ is a
complex $2 \times 2$ matrix.  Dropping the subscript, we
can always define
\beq
x=x^0 + x^a \s^a, \qquad x^\dagger = \xb^0 + \xb^a \s^a
\eeq
Then it is straightforward to work out ($\mu=0,\ldots,3$)
\beq
x x^\dagger = x^\mu \xb^\mu + (x^0 \xb^c + \xb^0 x^c
+ i x^a \xb^b \e^{abc} ) \s^c \equiv \phi^{x,0} + \phi^{x,c} \s^c
\equiv \phi^x
\eeq
Note that $\phi^{x,\mu}$ are real, and
that $\phi^{x,0}$ is positive definite.  With
similar definitions for $\phi^y, \phi^z$, the CKKU deformation is
\beq
\SSB &=& \frac{a^2 \mu^2}{2 g^2} \sum_\mbf \tr \[ \(\phi_\mbf^x-\frac{1}{2a^2}\)^2
+ \(\phi_\mbf^y-\frac{1}{2a^2}\)^2 + \frac{2}{a^2} \phi_\mbf^z \] \nnn
&=& \frac{a^2 \mu^2}{g^2} \sum_\mbf \[ \( \phi_\mbf^{x,0} - \frac{1}{2a^2}\)^2
+ \( \phi_\mbf^{y,0} - \frac{1}{2a^2}\)^2 
+ \frac{2}{a^2} \phi_\mbf^{z,0} \right. \nnn 
&& \quad \left. + \sum_a \[ (\phi_\mbf^{x,a})^2 + (\phi_\mbf^{y,a})^2 \] \]
\eeq
It can be seen that the deformation drives $\phi_\mbf^{x,a},\phi_\mbf^{y,a},
\phi_\mbf^{z,0}$ toward the origin, and $\phi_\mbf^{x,0}, \phi_\mbf^{y,0}$ 
toward $1/2a^2$.  When $\phi_\mbf^{z,0}=0$, 
it is easy to see that $\phi_\mbf^{z,a}=0$ identically.

To continue the analysis, 
it is convenient to rescale to dimensionless quantities using
the parameter $a$:
\beq
\hat g = g a^2, \quad \hat \mu = \mu a, \quad \hat \phi_\mbf^x = a^2 \phi_\mbf^x,
\quad \hx_\mbf = a x_\mbf, \quad {\rm etc.}
\label{krkr}
\eeq
Then
\beq
\SSB &=& \frac{{\hat \mu}^2}{{\hat g}^2} 
\sum_\mbf \[ \( \hphi_\mbf^{x,0} - \frac{1}{2}\)^2
+ \( \hphi_\mbf^{y,0} - \frac{1}{2}\)^2 
+ 2 \hphi_\mbf^{z,0}
+ \sum_a \[ (\hphi_\mbf^{x,a})^2 
+ (\hphi_\mbf^{y,a})^2 \] \]
\label{pool}
\eeq
For any value of the lattice spacing $a$, the minimum of $\SSB$ is
obtained iff
\beq
\hphi_\mbf^{x,0} = \hphi_\mbf^{y,0}= \half, \quad
\hphi_\mbf^{z,0} = \hphi_\mbf^{x,a} =\hphi_\mbf^{y,a} = 0,
\quad \forall \mbf
\label{jjus}
\eeq

The conditions involving $x_\mbf$ are just
\beq
\hx_\mbf^\mu \hxb_\mbf^\mu = \half, \qquad
\hx_\mbf^0 \hxb_\mbf^c + \hxb_\mbf^0 \hx_\mbf^c 
+ i \hx_\mbf^a \hxb_\mbf^b \e^{abc} = 0
\label{piwr}
\eeq
Let us examine what additional constraint this places
on classical solutions to $S=0$, beyond the restrictions
of the undeformed theory.  

First we note that neither of the non-zeromode branches
discussed in Section \ref{jirr} above are minima of
$\SSB$.  Thus we pass on to the zeromode configurations \myref{uier}.
Eqs.~\myref{piwr} then imply
that \myref{uier} is restricted to the form
\beq
\hx = \frac{e^{i \gamma_x}}{\sqtw} \diag \( e^{i \varphi_x},e^{-i \varphi_x} \)
\label{xsom}
\eeq
and global gauge transformations of this.
That is, $\hx$ is restricted to be an element of the 
maximal abelian subgroup $U(1)^2$ of $U(2)$, up to an
overall factor of $1/\sqtw$.  Similarly, we have for $\hy$,
\beq
\hy = \frac{e^{i \gamma_y}}{\sqtw} \diag \( e^{i \varphi_y},e^{-i \varphi_y} \)
\label{ysom}
\eeq
Finally, $\hphi_\mbf^{z,0} = 0$ implies $z_\mbf^\mu \zb_\mbf^\mu = 0$,
which has the unique solution $z_\mbf=0, \; \forall \mbf$.

Apart from global obstructions that are essentially Polyakov loops
in the $\ibf$ or $\jbf$ directions, the configuration \myref{xsom}
and \myref{ysom} can be gauged away.  It is straightforward to
verify that the required gauge transformation is \myref{yuer} with
\beq
\alpha_{m_1,m_2} = e^{i (m_1 \gamma_x + m_2 \gamma_y)} 
\times \diag \( e^{i(m_1 \varphi_x + m_2 \varphi_y)},
 e^{- i(m_1 \varphi_x + m_2 \varphi_y)} \)
\eeq
This sets all $\hx_\mbf,\hy_\mbf$ to unity except at the ``boundaries'':
\beq
\hx_{N,m_2} &=& \frac{e^{i N \gamma_x}}{\sqtw} 
\diag \( e^{i N \varphi_x},e^{-i N\varphi_x} \), \quad \forall m_2 \nnn
\hy_{m_1,N} &=& \frac{e^{i N \gamma_y}}{\sqtw} 
\diag \( e^{i N \varphi_y},e^{-i N\varphi_y} \), \quad \forall m_1
\eeq
For most purposes, we do not expect such vacua to distinguish
themselves from the trivial vacua in the thermodynamic limit.
In any case, global features such as these are typical of classical
vacua of other lattice Yang-Mills formulations, such as the Wilson
action.  In our simulation study we will avoid this issue by
restricting our attention to the expectation value of
a quantity that is independent of these angles.

\mys{Simulation}
\label{s:sim}
The emergence of the effective lattice theory relies upon
the assumption that fluctuations about this classical minimum
are small, and that the equations \myref{jjus}
are a good approximation to the corresponding expectation values
in the quantum theory.  For this reason we study
\beq
\vev{\hphi_\mbf^{x,0}} = \vev{\hx_\mbf^\mu \hxb_\mbf^\mu}
= \bigvev{ \half \tr (\hx_\mbf \hxd_\mbf) }
\label{iwre}
\eeq
in our simulations, and compare the
expectation values to the classical prediction \myref{jjus}.

\subsection{Scaling}
We study \myref{iwre} along a naive scaling trajectory:
\beq
g_2 = a^{-1} \hat g(a) = \mbox{fixed}
\label{yuie}
\eeq
That is, we hold the bare coupling in physical units, $g_2$, fixed;
this is equivalent to neglecting its anomalous dimension.  The dimensionless
bare coupling $\hat g$ is then a function of $a$ that vanishes
linearly with $a$ as the UV cutoff is removed.  

With regard to $\hmu$ we follow the instructions of CKKU:  
we send the dimensionless coefficient $\hat \mu$ of the deformation $\SSB$ to
zero as $1/N$ while increasing $N$.  
\beq
\hat \mu^{-1} = c N,  \qquad c = \ord{1}
\label{pwer}
\eeq
This is equivalent to scaling 
$\mu = 1/cL$, where $c$ is a constant and $L=Na$ is the
extent of the system.  

In the rescaled variables \myref{krkr},
the coefficient of the undeformed action is $1/\hg^2$,
whereas the coefficient of the deformation is $\hmu^2/ \hg^2$,
as can be seen from \myref{pool}.  Thus it is that
the relative strength of the deformation vanishes
in the thermodynamic limit, when \myref{pwer} is imposed.

We perform these scalings for a sequence of decreasing
values of $a$.  That is, we
study the thermodynamic limit for fixed values of $a$.  We then
extrapolate toward $a=0$ to obtain the continuum limit.

The physical length scales are set by 
$g_2^{-1}$ and the system size $L=Na$.
To keep discretization effects to a minimum we would like to
take $g_2^{-1} \gg a$.  Equivalently, $\hg^{-1} \gg 1$.  On the other
hand we are most interested in what happens at large or infinite
volume.  To render finite volume effects negligible would
require $g_2^{-1} \ll L$.  Equivalently, $\hg^{-1} \ll N$.
In the simulations we study the system for 
various choices of $\hg^{-1}$ and $N$.
We extrapolate to the regime $1 \ll \hg^{-1} \ll N$, but
often violate the bounds $1 \leq \hg^{-1} \leq N$ for specific
points where measurements are taken.  The reason for this
is that data outside the optimal window $1 \ll \hg^{-1} \ll N$
is informative to the extrapolation.

\subsection{Sampling procedure}
We update the system using a multi-hit Metropolis algorithm.
We attempt to update a single site or link
field 10 times before moving to the next, with an acceptance
rate of approximately 50 percent for a single hit.  We find
that this minimizes autocorrelations while maintaining program
efficiency.  We have examined autocorrelations and the
dependence of our observables on initial conditions.  
These studies have led us to make
500 themalization sweeps after a random initialization,
and 100 updating sweeps between each sample.  1/2 to 2 percent
standard errors result from accumulating 1000 samples at
each data point.

\subsection{Results}
\begin{figure}
\begin{center}
\includegraphics[height=5.0in,width=5.0in,angle=90]{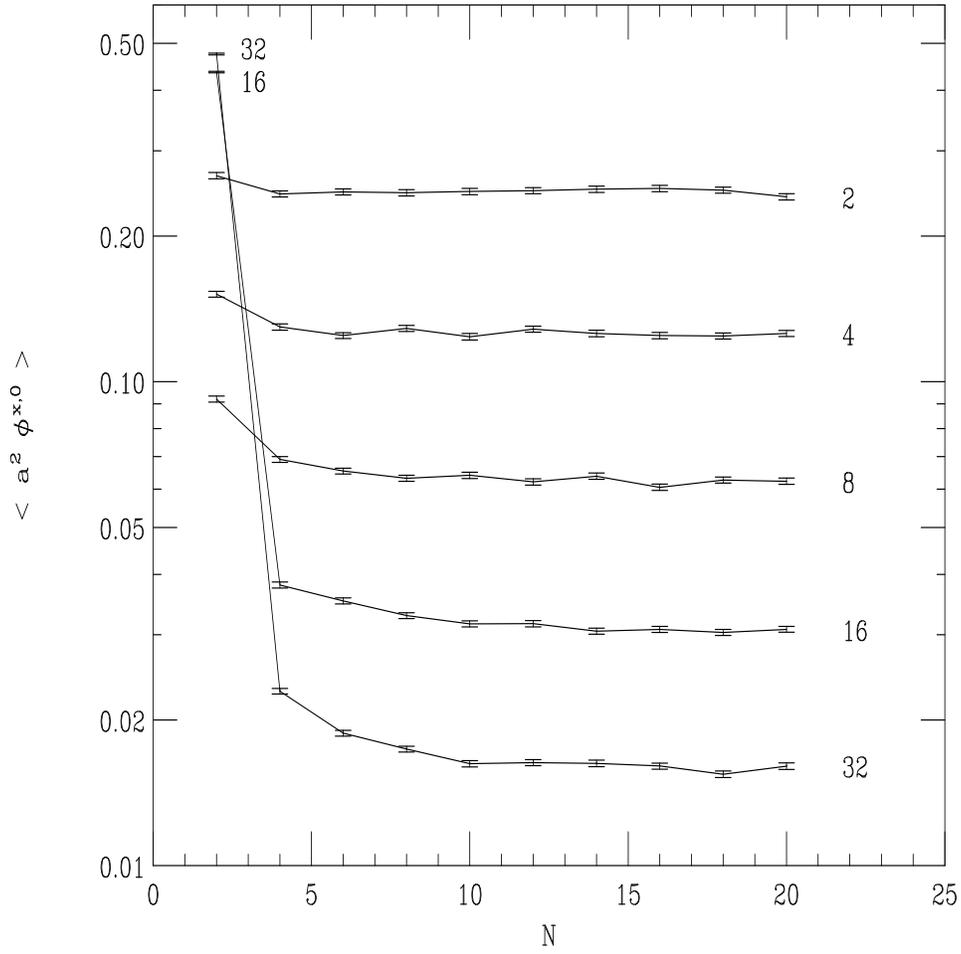}
\end{center}
\caption{Trajectories of fixed lattice spacing,
increasing volume, with data connected by
lines to guide the eye.  Each line is marked
by the corresponding value of $\hg^{-1}$.
A doubling of $\hg^{-1}$ corresponds
to a halving of the lattice spacing.}
\label{f1}
\end{figure}

In Fig.~\ref{f1} we show $\vev{\hphi_\mbf^{x,0}}$ [cf.~\myref{iwre}]
as a function of $N$
for various values of $\hg^{-1}$, having set $c=1$
in \myref{pwer}.  Doubling $\hg^{-1}$
is equated with halving the lattice spacing, according to \myref{yuie}.
It can be seen that $\vev{\hphi_\mbf^{x,0}}$ appears to tend toward
smaller values as the lattice regulator is removed, contrary to
the classical expectations \myref{jjus}.  At large enough values of $N$
the curves flatten out to a constant value. 

\begin{figure}
\begin{center}
\includegraphics[height=5.0in,width=2.5in,angle=90]{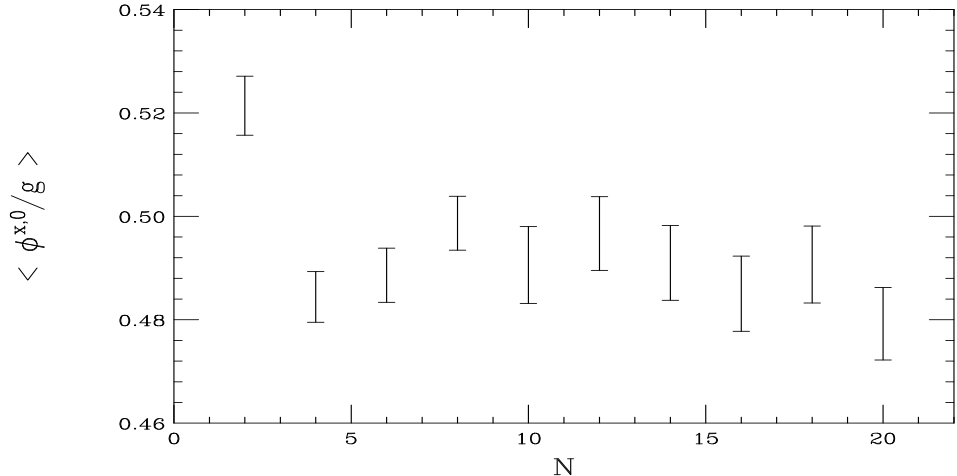}
\end{center}
\caption{Undeformed theory results.
\label{f2}}
\end{figure}

To understand this behavior, we first note that we are computing
an expectation value that is already nonvanishing in the undeformed
theory ($\mu \equiv 0$).  The undeformed theory expectation values 
\beq
\vev{g^{-1} \phi_\mbf^{x,0}}=
\vev{\hg^{-1} \hphi_\mbf^{x,0}}
\label{uter}
\eeq
for various values of $N$ are shown in Fig.~\ref{f2}.  The
rescaling by $g^{-1} = a^2 \hg^{-1}$ is useful because it corresponds to
removing $g$ from the undeformed action $S_0$ 
by a rescaling of the lattice variables [cf.~\myref{sbta}]; 
this amounts to studying the undeformed theory in units of $\sqrt{g}$.
It can be seen from Fig.~\ref{f2} that this quantity is far
from zero, and is rather insensitive to $N$.  The undeformed
theory only contains two length scales, $1/\sqrt{g}$ and $N/\sqrt{g}$.
For large $N$, it is not surprising that the {\it local} expectation
value \myref{uter} is insensitive to this long distance scale.  Rather,
it is determined by the short distance scale $1/\sqrt{g}$.
The deformed theory retains this short distance scale.  Once the
system size is much larger than this scale, the local expectation
value becomes independent of the system volume; this is particularly
true because the relative strength of the deformation is being scaled away
to zero [cf.~\myref{pwer}].

In Table \ref{tb2} we show the large $N$
expectation values \myref{uter} in the deformed theory
as well as those of the undeformed theory.
Here, the values for $N=16, 18, 20$
were averaged for each $\hginv$, which should provide
a good estimate of the asymptotic value, as can be
seen in Fig.~\ref{f1}.  The error was estimated
based on the maximum deviation from this mean, among
the three data points, taking into account the $1\s$ error
estimates that have been represented in the figure
by error bars.  Table \ref{tb2} shows that the large $N$
expectation values \myref{uter} {\it in the deformed theory}
are (up to statistical errors) the same as those of the 
{\it undeformed theory.}

\begin{table}
\begin{center}
\begin{tabular}{ccc}
$\hginv$ & $\vev{\hat \phi^{x,0}}$ & $\vev{\hg^{-1} \hphi^{x,0}}$ \\ \hline
2 & 0.2469(94) & 0.494(19) \\
4 & 0.1249(27) & 0.500(11) \\
8 & 0.0618(22) & 0.494(18) \\
16 & 0.03064(75) & 0.490(12) \\
32 & 0.01586(64) & 0.507(21) \\ \hline
1 ($\mu \equiv 0$) & & 0.496(22) \\ \hline
\end{tabular}
\caption{Large $N$ asymptotic values for the $\muinv=N$ trajectories.
For comparison, the result for the undeformed ($\mu \equiv 0$) expectation value
is shown in the bottom line.  Estimated errors in the last 2 digits
are shown in parentheses.
\label{tb2}}
\end{center}
\end{table}

\mys{Interpretation}
\label{intp}
In our simulations, we have observed that under the 
scaling \myref{pwer}, the deformation becomes ineffective
at changing the expectation value \myref{uter}, or
equivalently \myref{iwre}, away from the value that
would be obtained in the undeformed theory.  What has
happened is that the flat directions that were
lifted by the deformation are becoming flat all over
again as $N \to \infty$.  More precisely, the deformation
is proportional to $1/g_2^2 L^2$, and in the thermodynamic
limit this quantity vanishes.  Whereas
the configurations of the moduli space of the undeformed theory
that were lifted by $\SSB$ cost energy at finite $N$, 
the number of such configurations is
becoming vast due to the approximate flatness
in those directions.  For $N \gappeq \hg^{-1/2}$,
the entropy of these configurations wins out over
the energy arguments that prefer \myref{jjus}.

One lifted region of the $S_0$ moduli space 
where this is particularly clear is the branch \myref{kerr}.
The action for such configurations is (setting $c=1$)
\beq
S=\frac{1}{2 \hg^2} + \frac{2}{\hg^2 N^2} \sum_\mbf \(
|\hz_\mbf^0|^2 + |\hz_\mbf^3|^2 \)
= \frac{N^2}{2 g_2^2 L^2} + \frac{2}{g_2^2 L^2} \sum_\mbf \(
|\hz_\mbf^0|^2 + |\hz_\mbf^3|^2 \)
\eeq
Integrating $\exp(-S)$ over all $\hz_\mbf^{0,3}$ we obtain
\beq
\(\frac{\pi}{2}\)^{2N^2} 
\exp\[ \frac{N^2}{2 g_2^2 L^2} \( -1 + 4 g_2^2 L^2 \ln (g_2^2 L^2) \) \]
\eeq
Note that for large system size $g_2 L \gg 1$.  Thus, the positive
(entropic) term under the exponential wins out over the
negative (energetic) term by a large margin.  It would seem that
the weight of these configurations increases exponentially
as we increase $N$ while holding $g_2 L$ fixed; that is,
in the continuum limit.

Indeed the observed behavior summarized in Table~\ref{tb2}
is that in the deformed theory
\beq 
\bigvev{ \half \tr (\hx_\mbf \hxd_\mbf) } \approx \frac{g_2 L}{2N}
= \half g_2 a
\eeq
Thus the simulations likewise indicate that configurations
with $\hx_\mbf=0$ are dominating as we increase $N$ while
holding $g_2 L$ fixed, which is nothing but the continuum
limit.  By symmetry, the same also holds for $\hy_\mbf$.

\mys{Conclusions}
\label{s:con}
Naturally we find our results disappointing.  Attempts to formulate
supersymmetric field theories on the lattice have a long and troubled
history (see for example 
\cite{Curci:1986sm,Montvay:1994ze,Fleming:2000fa,Montvay:2001aj,Feo:2002yi} 
and references therein).  Recent success in non-gauge models
is very encouraging 
\cite{Catterall:2000rv,Catterall:2001wx,Catterall:2001fr}.  
Of particular importance is the understanding
of these models as ``topological'' or {\it Q-exact,} where $Q$ is an
exact supercharge of the lattice system 
\cite{Catterall:2003wd,Catterall:2003xx,Catterall:2003uf}.
Indeed, the CKKU undeformed action can be written in a Q-exact form.
An exciting development has been Sugino's exploitation of this Q-exact 
idea to constuct lattice super-Yang-Mills with compact gauge fields and
an ordinary (non-dynamical) lattice spacing \cite{Sugino:2003yb}.  
While the Sugino construction has its benefits, he notes
(based on a remark by Y. Shamir)
that these systems also suffer from a vacuum degeneracy
problem that renders the classical continuum limit
ambiguous.  Consequently, Sugino has introduced a non-supersymmetric
deformation in these models that lifts the unwanted
vacua; he demands that the relative strength of
this deformation be scaled to zero in the thermodynamic
limit.  Thus in some respects the actions of Sugino
suffer from a problem that is similar to that of CKKU.
However, he argues that entropic effects do not destroy the
vacuum selection imposed by his deformation.  We are
currently investigating this matter \cite{enm}.  

However, we continue to find the constructions of CKKU intriguing.
We would like to explore other possibilities for the
deformation.  In the non-supersymmetric case, we do not
see any compelling reason to scale the relative strength of the
deformation to zero in the thermodynamic limit.  We are
currently investigating whether or not a well-defined
continuum limit can be obtained for the present system
if this is not done.  Furthermore, we would like to better
understand the phase structure of the deformed theory.
Finally, it would be interesting to extend the present
analysis to the supersymmetric system of CKKU, as well as higher
dimensional systems.  In the former case we expect
supersymmetric nonrenormalization theorems to play
a role in the behavior of expectation values as the
deformation is scaled away.  In the latter case, spontaneous
symmetry breaking is allowed---since we do not face the
theorems special to 2d \cite{Mermin:1966fe,Coleman:ci}; 
so a nontrivial phase structure
may exist that would allow for a well-defined continuum
limit.

\vspace{15pt}

\noindent
{\bf \Large Acknowledgements}

\vspace{5pt}

\noindent
The author would like to thank Erich Poppitz for comments,
especially in regard to the moduli space of the undeformed
theory.
This work was supported by the National Science and Engineering 
Research Council of Canada and the Ontario 
Premier's Research Excellence Award.

\newpage

\myappendix

\mys{$\Ncal=4$ moduli space}
Here we establish the well-known solution to \myref{qers}.
One way to see this is as follows \cite{Fayet:1978ig}.  First we note that
$S_0$ reduced to the zero modes, which we write as $S_z$,
takes the form
\beq
S_z &=& \frac{N^2}{g^2} \tr \( \half ([\xdag,x] + [\ydag,y] + [\zdag,z])^2 \right.
\nnn && \left. +2 [x,y][\ydag,\xdag] + 2[y,z][\zdag,\ydag] +2 [z,x][\xdag,\zdag] \)
\label{jhas}
\eeq
Now note that the $U(1)$ parts of $x,y,z$ do not appear and can
take any value.  Thus we can restrict our attention the the $SU(k)$
parts, which we choose to express in terms of Hermitian matrices
$a_p,b_p, \; p=1,2,3$:
\beq
x^c T^c &=& (a_1^c + i b_1^c) T^c = a_1 + i b_1 \nnn
y^c T^c &=& (a_2^c + i b_2^c) T^c = a_2 + i b_2 \nnn
z^c T^c &=& (a_3^c + i b_3^c) T^c = a_3 + i b_3
\label{ksfe}
\eeq
Substitution into \myref{jhas} and a bit of algebra yields
\beq
S_z &=& -\frac{N^2}{g^2} \tr \[ 2 \(\sum_p [a_p,b_p]\)^2
+ \sum_{p,q} \( [a_p, b_q] + [b_p, a_q] \)^2 \right. \nnn
&& \left. + \sum_{p,q} \( [a_p, a_q] - [b_p, b_q] \)^2 \] \nnn
&=& -\frac{N^2}{g^2} \sum_{p,q} \tr \( [a_p, a_q]^2
+ [b_p, b_q]^2 + 2 [a_p, b_q]^2 \)
\eeq
Using positivity arguments quite similar to those above,
one finds that $S_z \geq 0$  and that $S_z = 0$ iff
\beq
[a_p, a_q] = [b_p, b_q] = [a_p, b_q] = 0 , \quad \forall p,q
\label{oiwe}
\eeq
which is nothing other than Eq.~(53) of \cite{Fayet:1978ig}.
Since the matrices are all hermitian and they all commute, 
it is obviously possible to choose a basis which simultaneously
diagonalizes them.  This basis will be related
to the one used in \myref{ksfe} according to 
$T^c \to T'^c = \alpha T^c \alpha^\dagger$,
which is nothing other than the global 
gauge transformations \myref{oore}.

\end{document}